\documentstyle[prl,aps,twocolumn]{revtex}

\input{psfig}
\begin{document}
\draft
\fnsymbol{footnote}

\wideabs{

\title{Gravitational Radiation Instability in Hot Young Neutron Stars
}

\author{Lee Lindblom and Benjamin J. Owen}
\address{Theoretical Astrophysics 130-33,
         California Institute of Technology,
         Pasadena, CA 91125}

\author{Sharon M. Morsink}
\address{Physics Department, 
         University of Wisconsin-Milwaukee,
         P.O. Box 413,
         Milwaukee, WI 53201}        

\date{\today}
\maketitle

\begin{abstract} We show that gravitational radiation drives an
instability in hot young rapidly rotating neutron stars.  This
instability occurs primarily in the $l=2$ {\it r}-mode and will
carry away most of the angular momentum of a rapidly rotating star
by gravitational radiation.  On the timescale needed to
cool a young neutron star to about $T=10^9$K (about one year) this
instability can reduce the rotation rate of a rapidly rotating star
to about $0.076\Omega_K$, where $\Omega_K$ is the Keplerian angular
velocity where mass shedding occurs.  In older colder neutron stars
this instability is suppressed by viscous effects, allowing older
stars to be spun up by accretion to larger angular velocities.

\end{abstract}
}

\narrowtext

Recently Andersson \cite{andersson} discovered (and Friedman and
Morsink \cite{friedman-morsink} confirmed more generally) that
gravitational radiation tends to drive the {\it r-}modes of all
rotating stars unstable.  In this paper we examine the timescales
associated with this instability in some detail.  We show that
gravitational radiation couples to these modes primarily through the
current multipoles, rather than the usual mass multipoles.  We also
evaluate the effects of internal fluid dissipation which tends to
suppress this instability.  We find that gravitational radiation is
stronger than viscosity in these modes and so this instability
severely limits the rotation rates of hot young neutron stars.  We
show that such stars can spin down by the emission of gravitational
radiation to about 7.6\% of their maximum rotation rates on the
timescale (about one year) needed to cool these stars to $10^9$K.
 
The {\it r-}modes of rotating barotropic Newtonian stars are solutions
of the perturbed fluid equations having (Eulerian) velocity
perturbations

\begin{equation}
\delta \vec{v} = \alpha R \Omega \left({r\over R}\right)^l \vec{Y}^{B}_{l\,l}
e^{i\omega t},\label{1}
\end{equation}

\noindent where $R$ and $\Omega$ are the radius and angular velocity of
the unperturbed star, $\alpha$ is an arbitrary constant, and
$\vec{Y}^{B}_{l\,m}$ is the magnetic type vector spherical harmonic
defined by

\begin{equation}
\vec{Y}^{B}_{l\,m}= [l(l+1)]^{-1/2}r\vec{\nabla}\times(r \vec{\nabla}Y_{l\,m}).
\label{2}
\end{equation}
 
\noindent Papaloizou and Pringle \cite{papaloizou-pringle} first
showed that the Euler equation for $r-$modes determines the frequencies as

\begin{equation}
\omega = - {(l-1)(l+2)\over l+1}\Omega.\label{3}
\end{equation}

\noindent Further use of the Euler equation (as first noted by
Provost, Berthomieu and Rocca \cite{provost}) in the barotropic case
(a good approximation for neutron stars) determines
that only the $l=m$ {\it r-}modes exist, and that $\delta \vec{v}$
must have the radial dependence given in Eq.~(\ref{1}).  These
expressions for the velocity perturbation and frequency are only the
lowest order terms in expansions for these quantities in powers of
$\Omega$.  The exact expressions contain additional terms of order
$\Omega^3$.

The lowest order expressions for the (Eulerian) density perturbation
$\delta \rho$ can also be deduced from the perturbed fluid equations
(Ipser and Lindblom \cite{ipser-lind}):

\begin{eqnarray}
{\delta \rho\over\rho} &=&
\alpha R^2\Omega^2 {d\rho\over dp} \nonumber \\
&&\times\left[{2 l\over 2l+1}\sqrt{l\over l+1}
\left({r\over R}\right)^{l+1}+\delta\Psi(r)\right] Y_{l+1\,l}
\,e^{i\omega t},\label{4}
\end{eqnarray}

\noindent where $\delta\Psi(r)$ is proportional to the gravitational
potential $\delta \Phi$ and satisfies

\begin{eqnarray}
{d^2\delta \Psi\over dr^2} &+& {2\over r} {d\delta \Psi\over dr}
+\left[4\pi G\rho {d\rho\over dp}- {(l+1)(l+2)\over r^2}\right]\delta \Psi
\nonumber\\&&
= -  {8\pi G l\over 2l+1} \sqrt{l\over l+1}
\rho {d\rho\over dp}\left({r\over R}\right)^{l+1}\!\!\!.\label{5}
\end{eqnarray}

\noindent Eq.~(\ref{4}) is the
complete expression for $\delta\rho$ to order $\Omega^2$. The
next order terms are proportional to $\Omega^4$. 

Our interest here is to study the evolution of these modes due to the
dissipative influences of viscosity and gravitational radiation.  For
this purpose it is useful to consider the effects of radiation on
the evolution of the energy of the mode (as measured in the co-rotating
frame of the equilibrium star) $\tilde{E}$:

\begin{equation}
\tilde{E}={1\over 2} \int\left[ \rho\, \delta \vec{v}\cdot \delta\vec{v}\,^*
+ \left({\delta p\over\rho} - \delta\Phi\right)  \delta \rho^*\right]d^3x.
\label{6}
\end{equation}

\noindent This energy evolves on the secular timescale of the
dissipative processes.  The general expression for the time derivative
of $\tilde{E}$ for a mode with time dependence $e^{i\omega t}$ and
azimuthal angular dependence $e^{im\varphi}$ is

\begin{eqnarray}
{d \tilde{E}\over dt} 
&=&-\int\left(2\eta\delta\sigma^{ab}\delta\sigma_{ab}^*
+\zeta\delta\sigma \delta\sigma^*\right)d^3x \nonumber\\
&&
 -\omega(\omega+m\Omega)\sum_{l\geq 2} N_l \omega^{2l}
\left(|\delta D_{lm}|^2+|\delta J_{lm}|^2\right).
\label{7}
\end{eqnarray}

\noindent The thermodynamic functions $\eta$ and $\zeta$ that appear in
Eq.~(\ref{7}) are the shear and bulk viscosities of the fluid.  The
viscous forces are driven by the shear $\delta \sigma_{ab}$ and
expansion $\delta\sigma$ of the perturbation,
defined by the usual expressions

\begin{equation}
\delta \sigma_{ab}={\scriptstyle {1\over 2}}
(\nabla_{\!a}\delta v_b+\nabla_{\!b}\delta v_a
-{\scriptstyle {2\over 3}}\delta_{ab}\nabla_{\!c}\delta v^c),
\label{8}
\end{equation}

\begin{equation}
\delta\sigma=\nabla_{\!a}\delta v^a.\label{9}
\end{equation}

\noindent Gravitational radiation couples to the
evolution of the mode through the mass $\delta D_{lm}$ and current $\delta
J_{lm}$ multipole moments of the perturbed fluid,

\begin{equation}
\delta D_{lm} = \int \delta\rho\, r^l Y^{*}_{l\,m} d^3x,\label{10}
\end{equation}

\begin{equation}
\delta J_{lm}= {2\over c}\sqrt{l\over l+1}
\int r^l (\rho \,\delta\vec{v}+\delta \rho\, \vec{v})\cdot
\vec{Y}^{B*}_{l\,m} d^3x,\label{11}
\end{equation}

\noindent with coupling constant

\begin{equation}
N_l = {4\pi G\over c^{2l+1}} {(l+1)(l+2)\over l(l-1)[(2l+1)!!]^2}.
\label{12}
\end{equation}

\noindent The terms in the expression for $d\tilde{E}/dt$ due to
viscosity and the gravitational radiation generated by the mass
multipoles are well known \cite{ipser-lindblom91}.  The terms involving
the current multipole moments have been deduced from the general
expressions given by Thorne \cite{thorne}. 

We can now use Eq.~(\ref{7}) to evaluate the stability of the {\it
r-}modes.  Viscosity always tends to decrease the energy $\tilde{E}$,
while gravitational radiation may either increase or decrease
it.  The sum that appears in Eq.~(\ref{7}) is positive definite; thus
the effect of gravitational radiation is determined by the sign of
$\omega(\omega+m\Omega)$.  For {\it r-}modes this quantity is negative
definite:

\begin{equation}
\omega(\omega+l\Omega)= 
-{2(l-1)(l+2)\over (l+1)^2}\Omega^2<0.\label{13}
\end{equation}

\noindent Therefore gravitational radiation tends to increase the energy of
these modes.  For small angular velocities the energy $\tilde{E}$ is
positive definite: the positive term $|\delta\vec{v}|^2$ in
Eq.~(\ref{6}) (proportional to $\Omega^2$) dominates the
indefinite term $(\delta p/\rho - \delta \Phi)\delta\rho^*$
(proportional to $\Omega^4$).  Thus, gravitational radiation tends to
make {\it every} {\it r-}mode unstable in slowly rotating stars.  This
confirms the discovery of Andersson \cite{andersson} and the more
general arguments of Friedman and Morsink \cite{friedman-morsink}.  To
determine whether these modes are actually stable or unstable in
rotating neutron stars, therefore, we must evaluate the magnitudes of
all the dissipative terms in Eq.~(\ref{7}) and determine
which dominates.

Here we estimate the relative importance of these dissipative effects in
the small angular velocity limit using the lowest order expressions
for the $r-$mode $\delta \vec{v}$ and $\delta \rho$ given in
Eqs.~(\ref{1}) and (\ref{4}).  The lowest order expression for the
energy of the mode $\tilde E$ is

\begin{equation}
\tilde{E} = {\scriptstyle {1\over 2}}
\alpha^2 \Omega^2 R^{-2l+2}\int_0^R\rho\,r^{2l+2}dr
.\label{14}
\end{equation}

The lowest order contribution to the
gravitational radiation terms in the energy dissipation
comes entirely from the current multipole
moment $\delta J_{l\,l}$.  This term can be evaluated
to lowest order in $\Omega$ using Eqs.~(\ref{1}) and (\ref{11}):

\begin{equation}
\delta J_{l\,l}={2\alpha \Omega\over cR^{l-1}}\sqrt{l\over l+1}
\int_0^R \rho\,r^{2l+2} dr.\label{15}
\end{equation}

\noindent The other contributions from gravitational radiation to the
dissipation rate are all higher order in $\Omega$.  The mass multipole
moment contributions are higher order because {\it a}) the density
perturbation $\delta\rho$ from Eq.~(\ref{4}) is proportional to
$\Omega^2$ while the velocity perturbation $\delta \vec{v}$ is proportional
to $\Omega$; and {\it b}) the density perturbation $\delta\rho$ generates
gravitational radiation at order $2l+4$ in $\omega$ while
$\delta\vec{v}$ generates radiation at order $2l+2$.

The contribution of gravitational radiation to the imaginary part of the
frequency of the mode $1/\tau_{\scriptscriptstyle GR}$ can be computed
as follows,

\begin{equation}
{1\over \tau_{\scriptscriptstyle GR}} 
=- {1\over 2 \tilde{E}} 
\left({d\tilde{E}\over dt}\right)_{\scriptscriptstyle GR}.
\label{16}
\end{equation}

\noindent Using Eqs.~(\ref{14})--(\ref{16}) we obtain an explicit
expression for the gravitational radiation timescale associated with the
{\it r-}modes:

\begin{eqnarray}
{1\over \tau_{\scriptscriptstyle GR}} 
&=&-{32\pi G\Omega^{2l+2}\over c^{2l+3}}\nonumber \\
&&\times
{ (l-1)^{2l}\over [(2l+1)!!]^2}
\left({l+2\over l+1}\right)^{2l+2}
\int_0^R\rho\,r^{2l+2} dr.\label{17}
\end{eqnarray}

The time derivative of the energy due to viscous dissipation is driven
by the shear $\delta \sigma_{ab}$ and the expansion $\delta \sigma$ of
the velocity perturbation.  The shear can be evaluated using
Eqs.~(\ref{1}) and (\ref{8}) and its integral over the constant $r$
two-spheres performed in a straightforward calculation.  Using the
formulae for the viscous dissipation rate Eq.~(\ref{7}) and the
energy Eq.~(\ref{14}), we obtain the contribution of shear viscosity
to the imaginary part of the frequency of the mode,

\begin{equation}
{1\over \tau_{\scriptscriptstyle V}} 
= (l-1)(2l+1) \int_0^R \eta r^{2l} dr
\left( \int_0^R \rho\, r^{2l+2} dr\right)^{-1}.
\label{18}
\end{equation}

The expansion $\delta \sigma$, which drives the bulk viscosity
dissipation in the fluid, can be re-expressed in terms of the density
perturbation.  The perturbed mass conservation law gives the
relationship $\delta\sigma = -i(\omega+m\Omega)\Delta\rho/\rho$, where
$\Delta\rho$ is the Lagrangian perturbation in the density.  The
perturbation analysis used here is not of sufficiently high order (in
$\Omega$) to evaluate the lowest order contribution to $\Delta \rho$.
However, we are able to evaluate the Eulerian perturbation
$\delta\rho$ as given in Eq.~(\ref{4}).  We expect that the
integral of $|\delta \rho/\rho|^2$ over the interior of the star will
be similar to (i.e., within about a factor of two of) the integral of
$|\Delta \rho/\rho|^2$.  Thus, we estimate the magnitude of the
bulk viscosity contribution to the dissipation by

\begin{equation}
{1\over \tau_{\scriptscriptstyle B}}
\approx {(\omega+m\Omega)^2\over 2\tilde{E}}\int\zeta
{\delta\rho\,\delta\rho^*\over\rho^2}d^3x.\label{19}
\end{equation}

\noindent Using Eqs.~(\ref{4}) and (\ref{14}) for $\delta\rho/\rho$
and $\tilde{E}$, Eq.~(\ref{19}) becomes an explicit formula for the
contribution to the imaginary part of the frequency due to bulk
viscosity.

To evaluate the dissipative timescales associated with the {\it
r-}modes using the formulae in Eqs.~(\ref{17})--(\ref{19}), we need
models for the structures of neutron stars as well as
expressions for the viscosities of neutron star matter.  We have
evaluated these timescales for $1.4M_\odot$ neutron star
models based on several realistic equations of state \cite{paris}.
We use the standard formulae for the shear and bulk viscosities of hot
neutron star matter \cite{visrefs}

\begin{equation}
\eta=347\rho^{9/4}T^{-2},\label{20}
\end{equation}

\begin{equation}
\zeta = 6.0\times 10^{-59} \rho^2 (\omega+m\Omega)^{-2} T^6,
\label{21}
\end{equation}

\noindent where all quantities are given in cgs units.
The timescales for the more
realistic equations of state are comparable to those based on a simple
polytropic model $p=\kappa\rho^2$ with $\kappa$ chosen so that the
radius of a $1.4M_\odot$ star is 12.53 km.  The dissipation timescales
for this polytropic model (which can be evaluated analytically) are
$\tilde{\tau}_{\scriptscriptstyle GR}= -3.26$s,
$\tilde{\tau}_{\scriptscriptstyle V}=2.52\times 10^8$s and
$\tilde{\tau}_{\scriptscriptstyle B}=6.99\times 10^8$s for the
fiducial values of the angular velocity $\Omega=\sqrt{\pi G
\bar{\rho}}$ and temperature $T=10^9$K in the $l=2$ {\it r-}mode.  The
gravitational radiation timescales increase by about one order of
magnitude for each incremental increase in $l$, while the viscous
timescales decrease by about 20\%.

The evolution of an {\it r-}mode due to the dissipative effects
of viscosity and gravitational radiation reaction is determined
by the imaginary part of the frequency of the mode,

\begin{eqnarray}
{1\over \tau(\Omega)}
&=& {1\over  \tilde{\tau}_{\scriptscriptstyle GR}}
\left({\Omega^2\over \pi G \bar{\rho}}\right)^{l+1}
+ {1\over  \tilde{\tau}_{\scriptscriptstyle V}}
\left({10^9 {\rm K}\over T}\right)^2\nonumber\\
&&+ {1\over  \tilde{\tau}_{\scriptscriptstyle B}}
\left({T\over 10^9 {\rm K}}\right)^6
\left({\Omega^2\over \pi G \bar{\rho}}\right).
\label{22}
\end{eqnarray}

\noindent Eq.~(\ref{22}) is displayed in a form that makes explicit
the angular velocity and temperature dependences of the various terms.
Dissipative effects cause the mode to decay exponentially as
$e^{-t/\tau}$ (i.e., the mode is stable) as long as $\tau > 0$.
From Eqs.~(\ref{17})--(\ref{19}) we see that
$\tilde{\tau}_{\scriptscriptstyle V}>0$ and
$\tilde{\tau}_{\scriptscriptstyle B}>0$ while
$\tilde{\tau}_{\scriptscriptstyle GR}<0$.  Thus gravitational
radiation drives these modes towards instability while viscosity
tries to stabilize them.  For small $\Omega$ the gravitational
radiation contribution to the imaginary part of the frequency is very
small since it is proportional to $\Omega^{2l+2}$.  Thus for
sufficiently small angular velocities, viscosity dominates and the
mode is stable.  For sufficiently large $\Omega$, however,
gravitational radiation will dominate and drive
the mode unstable.  It is convenient to define a critical angular
velocity $\Omega_c$ where the sign of the imaginary part of the
frequency changes from positive to negative: $1/\tau(\Omega_c) = 0$.
If the angular velocity of the star exceeds $\Omega_c$ then
gravitational radiation reaction dominates viscosity and the mode
is unstable.

\begin{figure}
\centerline{\psfig{file=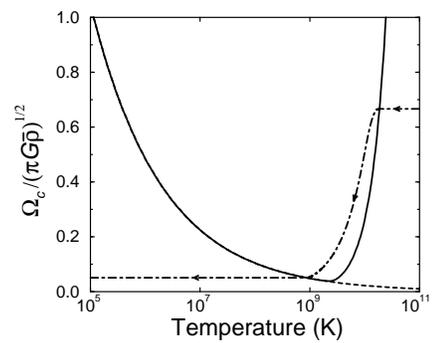,height=1.8in}}
\caption{Critical angular velocities for a $1.4M_\odot$ polytropic
neutron star with (solid) and without (dashed) bulk viscosity.  Also
the evolution of a rapidly rotating neutron star (dash-dot) as the
star cools and emits gravitational radiation.
\label{fig1}}
\end{figure}

\begin{figure}
\centerline{\psfig{file=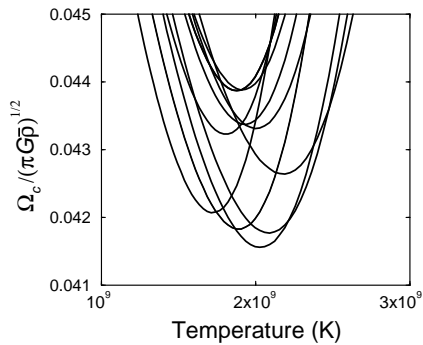,height=1.8in}}
\caption{Critical angular velocities of realistic $1.4M_\odot$ neutron
star models.\label{fig2}}
\end{figure}

For a given temperature and mode $l$ the equation for the critical angular
velocity, $0=1/\tau(\Omega_c)$, is a polynomial of order $l+1$ in
$\Omega_c^2$, and thus each mode has its own critical angular velocity.
However, only the smallest of these (always the $l=2$ {\it r-}mode here)
represents the critical angular velocity of the star.
Fig.~\ref{fig1} depicts the critical angular velocity for a range
of temperatures relevant for neutron stars.  The solid curve in
Fig.~\ref{fig1} represents the critical angular velocity for the
polytropic model discussed above.  Fig.~\ref{fig2} depicts
the critical angular velocities for $1.4M_\odot$
neutron star models computed from a variety of realistic equations of
state \cite{paris}.  Fig.~\ref{fig2} illustrates that the minimum
critical angular velocity (in units of $\sqrt{\pi G \bar{\rho}}$) is
extremely insensitive to the equation of state. The minima of these
curves occur at $T \approx 2\times 10^9$K, with
$\Omega_c \approx 0.043\sqrt{\pi G \bar{\rho}}$.  The maximum angular velocity
for any star occurs when the material at the surface effectively
orbits the star.  This `Keplerian' angular velocity $\Omega_K$ is very
nearly ${2\over 3}\sqrt{\pi G \bar{\rho}}$ for any equation of
state. Thus the minimum critical angular velocity due to instability
of the {\it r-}modes is about $0.065\Omega_K$ for any equation of
state \cite{footnote}.

\break

To determine how rapidly a young neutron star is allowed to spin after
cooling, we must compare the rate it cools with the rate it loses
angular momentum by gravitational radiation.  We approximate the
cooling with a simple model based on the emission of
neutrinos through the modified URCA process \cite{shapiro-teukolsky}.
We compute the time evolution of the angular velocity of the star by
setting $dJ/dt = J/\tau$, where $J$ is the angular momentum of the
star and $\tau$ is the timescale given in Eq.~(\ref{22}).  The result
is a simple first order differential equation for $\Omega(t)$ which we
solve for initial angular velocity
$\Omega=\Omega_K$ and initial temperature $10^{11}$K.  The solution is
shown as the dash-dot line in Fig.~\ref{fig1}.  The gravitational
radiation timescale is so short that the star radiates away its
angular momentum almost as quickly as it cools.  The angular
velocity of the star decreases from $\Omega_K$ to $0.076\Omega_K$ in a
period of about one year \cite{footnote2}.  Thus, we conclude that
young neutron stars will be spun down by the emission of gravitational
radiation within their first year to a rotation period
of about $13P_{\rm min}$, where $P_{\rm min}=2\pi/\Omega_K$.  The Crab
pulsar with present rotation period 33ms and initial period 19ms
(based on the measured braking index) rotates more slowly than this
limit if $P_{\rm min} < 1.5$ms.

Our analysis here is based on the assumption that a young hot neutron
star may be modeled as a simple ordinary fluid.  Once the star cools
below the superfluid transition temperature (about $10^9$K) the
analysis presented here must be modified \cite{footnote3}.  We expect
the {\it r-}mode instability to be completely suppressed (with
$\Omega_c=\Omega_K$) when the star becomes a superfluid
\cite{lindblom-mendell}.  This makes it possible for old recycled
pulsars to be spun up to large angular velocities by accretion if they
are not re-heated much above the Eddington temperature of $10^7$K in
the process.  If non-perfect fluid effects enter above $10^9$K,
however, the spin down process may be terminated at a higher angular
velocity than the $0.076\Omega_K$ figure computed here.  The detection
of a young fast pulsar would provide evidence for such non-perfect 
fluid effects at these high temperatures.

\acknowledgments

We thank N. Andersson, J. Friedman, J. Ipser, S. Phinney, B. Schutz,
and K. Thorne for helpful discussions.  This
research was supported by NSF grants AST-9417371, PHY-9507740,
PHY-9796079, NASA grant NAG5-4093, an NSF graduate fellowship, and
NSERC of Canada.



\begin{references}
\bibitem{footnote} Note that the minimum $\Omega_c$ is rather small,
and thus the small $\Omega$ expansions used here are expected to be
quite good.  Also, the approximation used to evaluate $\tau_B$ in
Eq.~(\ref{19}) is not expected have a large effect on this minimum
value: e.g., if $\tilde{\tau}_{\scriptscriptstyle B}$ were off by a
factor of ten, then the high $T$ portion of the curve in
Fig.~\ref{fig1} would be moved along the $T$ axis by a factor of about
$1.5\approx (10)^{1/6}$.  Since the shear viscosity influence on the
curve (the dashed line in Fig.~\ref{fig1}) is so flat in this
temperature range, the minimum value of $\Omega_c$ would not be
significantly changed.
\bibitem{footnote2} Our assumption that the spindown timescale is
$\tau$ of Eq.~(\ref{22}) is based on the fact that the $r-$mode will
grow (within a few minutes) to a point where the mode contains a
substantial fraction of the total angular momentum of the star before
being saturated by non-linear effects.  Eq.~(\ref{22}) is a low
$\Omega$ expansion so the early time values of $\tau$ are also
somewhat uncertain.  Fortunately the final angular velocity state of
the star is quite insensitive to the details of the cooling and
spindown.
\bibitem{footnote3} The formation of a solid crust below
$\rho\approx2\times10^{14}$g/cm$^3$ also affects these modes.  However,
our estimates suggest that this density is too low to substantially
affect the $r-$mode instability.
\end{references}
\end{document}